\documentclass[preprint2]{aastex}

\newcommand{\magn}{mag}
\newcommand{\jh}{\textit{J--H}}
\newcommand{\hk}{\textit{H--K}}
\newcommand{\ji}{\textit{J}}
\newcommand{\hi}{\textit{H}}
\newcommand{\ki}{\textit{K}}
\newcommand{\ksi}{\textit{K}\ensuremath{_{s}}}
\newcommand{\udagger}{\ensuremath{^{\dagger}}}
\newcommand{\chis}{\ensuremath{\chi^{2}}}
\newcommand{\nh}{\ensuremath{N_{\rm H}}}
\newcommand{\kt}{\ensuremath{k_{B}T}}
\newcommand{\lx}{\ensuremath{L_{X}}}
\newcommand{\Z}{\ensuremath{Z}}

\shorttitle{X-ray Properties of YSOs in OMC-2/3 from CXO}
\shortauthors{Tsujimoto et al.}

\begin{document}

\title{X-ray Properties of Young Stellar Objects in OMC-2 and OMC-3 from the \textit{Chandra} X-ray Observatory}
\author{Masahiro Tsujimoto and Katsuji Koyama}
\affil{Department of Physics, Graduate School of Science, Kyoto University, Sakyo-ku, Kyoto, 606-8502, Japan}
\email{tsujimot@cr.scphys.kyoto-u.ac.jp, koyama@cr.scphys.kyoto-u.ac.jp}
\author{Yohko Tsuboi}
\affil{Department of Physics \& Astrophysics, 525 Davey Laboratory, Pennsylvania State
University, University Park, PA 16802, USA}
\email{tsuboi@astro.psu.edu}
\and
\author{Miwa Goto and Naoto Kobayashi}
\affil{National Astronomical Observatory of Japan, 650 North A'ohoku Place, Hilo, HI 96720, USA}
\email{mgoto@subaru.naoj.org, naoto@subaru.naoj.org}

\begin{abstract}
We report X-ray results of the \textit{Chandra} observation of Orion Molecular Cloud 2
and 3. A deep exposure of $\sim$ 100~ksec detects $\sim$ 400 X-ray sources in the field
of view of the ACIS array, providing one of the largest X-ray catalogs in
a star forming region. Coherent studies of the source detection, time variability, and
energy spectra are performed. We classify the X-ray sources into class~I, class~II, and
class~III $+$ MS based on the \ji, \hi, and \ki-band colors of their near infrared
counterparts and discuss the X-ray properties (temperature, absorption, and time 
variability) along these evolutionary phases.
\end{abstract}


\section{INTRODUCTION}
Stars are known to possess high energy phenomena well before they reach the main
sequence (MS)\footnote{In this paper, we define terminologies as follows; ``Protostars''
are class 0 and class I objects. ``Pre-main sequence (PMS) stars'' and ``T Tauri
Stars (TTS)'' are class II and class III objects. TTS comprise two subclasses,
Classical T Tauri Stars (CTTS) and Weak-line T Tauri Stars (WTTS), corresponding to class
II and class III, respectively. ``Young Stellar Objects (YSO)'' are class 0, I, II, and
III sources collectively.}. The \textit{Einstein} observatory discovered that PMS stars (class III
and class II stage with ages of $\sim$ 10$^{7}$ and $\sim$ 10$^{6}$ years after the
on-set of gravitational collapse, respectively) are strong X-ray emitters
\citep{feigelson81, montmerle83}. Successive observations with \textit{ASCA} and
\textit{ROSAT} detected hundreds of X-ray samples of PMS stars with better spectral or
spatial resolution. They revealed that the X-ray activities in these objects are quite
common. \citet{koyama96} and \citet{kamata97} moved the start of X-ray activity forward to
protostars. They detected hard X-rays from class I ($\sim$ 10$^{5}$ years)
protostars with \textit{ASCA}. Recently, \citet{tsuboi01} reported hard X-ray emission
from two class 0 candidates in the Orion Molecular Cloud (OMC) using the \textit{Chandra}
X-ray Observatory (CXO). Now YSOs at virtually all the evolutionary phases --- from
class 0 to class III --- are known to emit X-rays.

However, the X-ray emission mechanism has not been well understood. Imanishi, Koyama,
\& Tsuboi~(2001) observed the $\rho$-Ophiuchi dark cloud with the CXO and found that the
X-ray emitting regions of several sources are well beyond their stellar surface. This may
be explained if the X-rays are due to star-disk magnetic interaction. \citet{grosso00}
reported, on the other hand, that the X-ray emission does not depend on the existence of
disks; no significant difference of the X-ray luminosity function is seen between
\textit{ROSAT}-detected class III objects (no disk) and class II objects (with
disks). \citet{montmerle00} advocated that the reconnection of magnetic loops between
stellar surface and disk is responsible for quasi-periodic flares and extremely high
luminosity X-rays found in a protostar (YLW16) with \textit{ASCA} \citep{tsuboi00}. 
\citet{schulz01} detected high temperature plasma with moderate luminosity and no rapid
variability from YSOs in the Orion Nebula Cluster with the CXO. Hence they argued that the
stellar surface-disk arcade model is unlikely but magnetically confined stellar plasma
is a more likely origin for the X-ray activities, as was suggested by \citet{skinner98}
based on the \textit{ASCA} observation of a CTTS (SU Aur). These arguments from previous
studies may not be disagreements with each other, but are far from a unified picture
of X-ray emission from YSOs.

Orion Molecular Cloud 2 and 3 (OMC-2 and OMC-3), located at a distance of $\sim$ 450~pc
from us, are the best sites to study X-ray properties of YSOs at various evolutionary
phases from class 0 through class III. X-rays from class 0 candidates have only been
reported in OMC-3 \citep{tsuboi01}. Therefore, this cloud is the unique site where we
actually provide X-ray samples from class 0 to class III. Moreover, the moderate angular
size allows us to fully cover the clouds in the CXO/ACIS field of view (FOV). 

Here, we report results of the CXO observation of the OMC-2 and OMC-3 star forming
regions, with one of the deepest exposures of star forming regions. The purposes of this
paper are 1)~to conduct a coherent study of the X-ray source detection, time
variability, and energy spectra, and to provide a catalog of one of the largest samples of
X-ray sources in a star forming region, and 2)~to investigate the X-ray properties of
YSOs as a function of evolutionary phase inferred from the near infrared (NIR) data, and
to compare them with the results of other star forming regions.

For the second purpose we use the 2MASS data, which is the deepest currently available
in these regions. We conducted deeper infrared surveys on these clouds down to $\sim$
18 \magn\ in the \ji, \hi, and \ki-band. Those results will be presented in a separate
paper.

\section{OBSERVATION}
The observation of OMC-2 and OMC-3 was carried out on 2000 January 1--2 using the CXO
with a nominal exposure time of 89.2~ksec. We used four ACIS-I chips (I0, I1, I2, and
I3) and one ACIS-S chip (S2) on the focal plane of the mirror system.
All these detector chips utilize front-illuminated CCDs, which have
sensitivity over a wide energy range (0.2--10.0~keV) with moderate energy
resolution ($\Delta E \sim$ 200~eV at $E=$ 6~keV). Together with the optics, ACIS achieves
sub-arcsec spatial resolution for on-axis sources, and a $8\farcm4 \times 8\farcm4$
FOV for each CCD chip. Details on the satellite and the detectors are found in
\citet{weisskopf96} and \citet{garmire01}, respectively.

Its spectroscopic capability in the hard energy band, together with the sub-arcsec
spatial resolution, makes the CXO/ACIS an ideal instrument for the observation of star
forming regions, where sources are heavily absorbed and crowded.

\begin{figure}
 \figurenum{1}
 \epsscale{1}
 \plotone{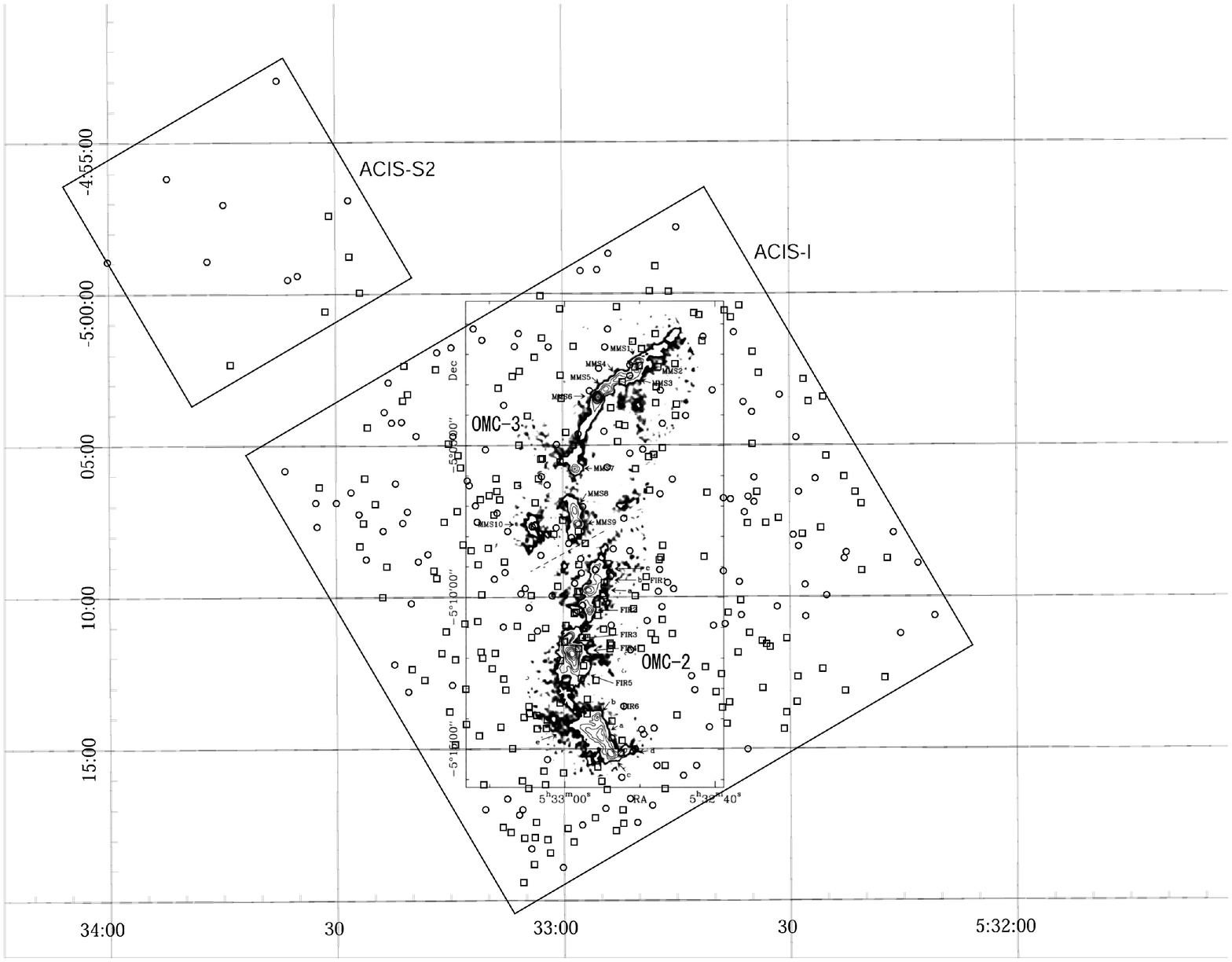}
 \caption{FOVs of ACIS-I (large square region) and ACIS-S2 (small square region at the
 top left) are overlaid on the 1300mm intensity map \citep{chini97}. ACIS-I covers the
 whole OMC-2 and OMC-3 star forming regions. Squares and circles are X-ray sources with
 and  without a 2MASS counterpart, respectively. The coordinate is in the equinox
 1950B.\label{fg:f1}}
\end{figure}

Figure~\ref{fg:f1} shows the FOV of our observation overlaid on the 1300mm
intensity contour map \citep{chini97}. OMC-2 and OMC-3 are separated into south and
north by the dotted line. Our observation fully covers both star forming clouds.

\placefigure{fg:f1}

\section{DATA REDUCTION AND ANALYSIS}
\subsection{Data Reduction}
For the data reduction, we use the level 2 data ``reprocessed'' at the Chandra X-ray
Center (CXC). This version improves the aspect solution and restores the degradation of
energy gain and resolution due to the increase of the charge transfer inefficiency (CTI)
of the ACIS\footnote{see http://asc.harvard.edu/udocs/reprocessing.html}. For data
manipulation, we use the Chandra Interactive Analysis of Observations (\texttt{CIAO})
version 2.1 and \texttt{FTOOLS} version 4.2. 

In order to avoid the ``spiking'' problem reported by the CXC\footnote{see
http://asc.harvard.edu/ciao/caveats/acis\_pi.html}, we manually randomize the values of
\texttt{ENERGY} and \texttt{PI} columns in the event file. After this procedure,
we conduct the analysis in the following sections separately for the ACIS-I and ACIS-S
data. Throughout this paper, we use X-ray photons in the 0.5--8.0~keV energy band. The
photons of each source are accumulated from an elliptical region. The major and minor
axes, and the rotation angles are derived from the \texttt{wavdetect} command. 

\subsection{Source Finding}
We use \texttt{wavdetect} command with the significance criterion of $1 \times 10^{-5}$
and the wavelet scales ranging from 1 to 16 pixels in multiples of $\sqrt{2}$. We remove
a few spurious sources through careful inspection by eye. We then detect 365 sources in
the 0.5--8.0~keV band image. In order to pick up either soft (less absorbed) or hard
(highly absorbed) sources more effectively, we also apply the same detection algorithm
to the 0.5--2.0~keV (soft) and 2.0--8.0~keV (hard) band image. Then 17 and 16 new
sources are found in the hard band and the soft band image, respectively. In total, we
detect 398 X-ray sources in the ACIS-I and ACIS-S FOV. For each detected source, we
calculate the X-ray photon count (0.5--8.0~keV) and the hardness ratio
(Table~\ref{tb:t1}). 

For a systematic study, we divide all the X-ray sources into two groups, ``bright''
($>$ 200 counts) and ``faint'' ($\leq$ 200 counts) sources, according to the photon
counts inside the accumulation region. Out of 398 sources, 136 sources are ``bright'',
and 262 are ``faint''. If the X-ray counts of the ``bright'' sources are less than 3
times the background counts in the photon accumulation region, we remove these
sources and define the remaining 123 sources to be ``bright2''. This screening is
practically necessary to comprise a good sample for spectral and timing analyses. In
particular, those at large off-axis angles have the background counts comparable to 200 due to
a rather large accumulation area, hence the background has a large impact on the quality of the source
spectrum and light curve.

\placetable{tb:t1}

\subsection{Correlation with 2MASS Sources}
For all the detected X-ray sources, we search for a NIR counterpart using the Point
Source Catalog in the 2MASS Second Incremental Data Release\footnote{see
http://www.ipac.caltech.edu/2mass/}. It covers the whole ACIS-I and ACIS-S FOVs, and provides
us with the NIR source positions and their photometric data in the \ji, \hi, and \ksi-band down to
15.8, 15.1, and 14.3 \magn, respectively.  

In the ACIS-I and ACIS-S FOVs, we find 638 2MASS sources.  First, we search for the 2MASS sources 
nearest to each CXO source within 3$^{\prime\prime}$ radius. Second, we conversely
search for the CXO source nearest to each 2MASS source within 3$^{\prime\prime}$ radius.
Thus we pick up the nearest CXO-2MASS pairs. The systematic position off-sets of the CXO
sources from their 2MASS counterparts is found to be $-0\farcs186$ and $0\farcs200$ in
the direction of right ascension and declination. After correcting these systematic
off-sets of the X-ray position, we re-apply the same procedure for the 2MASS counterpart
search. Finally we find that 238 out of 398 ($\sim$ 60\%) X-ray sources have a 2MASS
counterpart. The distance between X-ray sources and their counterparts is found to be
$\sim$ 0\farcs5.

Together with the X-ray properties, the \jh\ and \hk\ colors of their 2MASS
counterparts are given in Table~\ref{tb:t1}. About 80\% of the ``bright'' sources have a
2MASS counterpart (111 out of 136), while for the ``faint'' sources, the ratio is about 50\%
(127 out of 262).

\subsection{Timing Analysis} 
We make the X-ray light curve (background is not subtracted) for the ``bright2''
sources and perform \chis\ fit with a constant flux assumption. We discriminate the
time-variable sources by the significance criteria of 0.01, which are marked with
$\dagger$ in Table~\ref{tb:t2}. About 40\% (47 out of 123) of the sample sources are
found to be time-variable. Most of them show flare-like variability, a fast rise and
slow decay of the flux. Some light curves show multiple flares during the
observation. A variety of features are seen in the light curves, and it is difficult to
separate X-ray photons during the flare and at quiescence in a coherent manner. We
therefore deal with them in the same way.

\subsection{Spectral Analysis}
We next perform spectral analysis of the ``bright2'' sources, the same data set for the
timing analysis. We combine each energy bin so as to have more than 20 photons, then
subtract a background spectrum. We use the \texttt{sherpa} program for the spectral fitting.

First we fit the spectrum of all the sample sources with a thin thermal plasma and a
power-law model, both with absorption of hydrogen column density (\nh). The free
parameters of the former model are temperature (\kt), metallicity (\Z), and
normalization, while the latter's are photon index ($\Gamma$) and normalization. The
former model is accepted for 90 out of 123 sources with the upper probability of larger
than 0.01 (99\% confidence). The latter model is accepted for 64 sources, of which all
except two are also accepted in the former model. On the contrary, both the models are
rejected for the other 31 sources. Therefore, here and after, we use the results of the
thin-thermal plasma model fittings (Table~\ref{tb:t2}).

\placetable{tb:t2}

Second, for the 31 sources which reject both models, we try 2-component models, 1)~a
thin-thermal plasma $+$ power-law and 2)~2-temperature thin-thermal plasma
model. Additional free parameters are \kt\ (or $\Gamma$) and normalization. Both models
are acceptable for 8 and rejected for 6 sources out of 31 samples, while the latter
model are accepted by other 15 sources. We therefore use the results of the two temperature
thin-thermal plasma model fittings (Table~\ref{tb:t3}).

\placetable{tb:t3}

\section{DISCUSSION}
\subsection{The Nature of X-ray and NIR Sources}
\subsubsection{The Nature of NIR Sources}
Since some fraction of the 2MASS sources may be background or foreground sources, we
first estimate the contribution of background galaxies in the following way. The number
counts of galaxies per square degree per magnitude at a certain \ki-band (2.2\micron)
magnitude is given by
\begin{equation}
 \frac{dN}{dK} = 4000 \times 10^{\alpha (K-17)}
\end{equation}
where $\alpha = 0.67$ for 10 \magn $<K<$ 17 \magn\ \citep{tokunaga00}. Therefore, the number
in the range of $K_{min}$ \magn $<K<$ $K_{max}$ \magn\ is
\begin{equation}
 \int _{K_{min}}^{K_{max}} \frac{dN}{dK}\ dK= \frac{4000}{\alpha \log{10}} \{10^{\alpha(K_{max}-17)}-10^{\alpha(K_{min}-17)}\}.
  \label{eq:integrated}
\end{equation}
The NIR counterparts of the CXO sources have a \ki-band flux of 6 \magn\ $<K_{s}<$ 15
\magn. We substitute $K_{min}=6$ and $K_{max}=15$ for simplicity, though equation
(\ref{eq:integrated}) is valid only for $K_{min}>$10 \magn.  Still, this gives us a good
estimate, because the second term of the right hand side of equation
(\ref{eq:integrated}) is negligible compared to the first term in this case. Considering
our FOV ($8\farcm4 \times 8\farcm4$ for each chip), the estimated number of galaxies
with 6 \magn $<K_{s}<$ 15 \magn\ is $\sim$12. This is only 1.9\% of the 2MASS source
number in our FOV. In addition, the background galaxies in this direction may suffer
significant extinction due to molecular clouds, hence the contribution of extragalactic
sources to our 2MASS sources should be even smaller than that estimated above.

We can also neglect the contribution of foreground sources to our 2MASS sample, because
OMC-2 and OMC-3 lie off the galactic plane by $\sim$ 20 degrees. Thus, we can safely
assume most of the NIR sources to be cloud members (YSOs and MS stars).

\subsubsection{The Nature of X-ray Sources}
\citet{krishnamurthi01} observed the core of the Pleiades star cluster with the CXO for
36~ksec, and found that a significant fraction of X-ray sources are likely to be AGNs. We
therefore try to discriminate cloud members from extragalactic sources using the X-ray
hardness ratio.

\begin{figure}
 \figurenum{2}
 \epsscale{0.8}
 \plotone{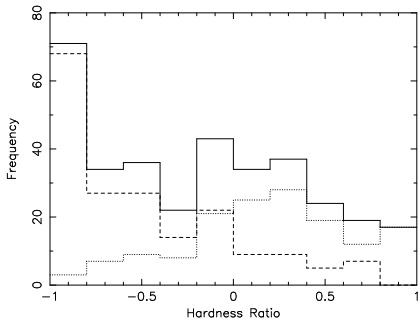}
 \caption{The histogram of the hardness ratio (from --1 to 1 with a step of 0.2) is
 shown by the dashed and dotted line for the X-ray sources with and without a NIR
 counterpart, respectively. The total is given by the solid line.\label{fg:f2}}
\end{figure}

In Figure~\ref{fg:f2}, the histogram of the hardness ratio is given separately for X-ray
sources with and without a NIR counterpart. The hardness ratio of the X-ray sources with
a NIR counterpart has the peak at $-1.0 \sim -0.8$, while those without a
NIR counterpart has its peak at $0.2 \sim 0.4$. A power-law spectrum of
$\Gamma=$1.7 has the hardness ratio of $0.1 \sim 0.4$ when absorbed with the column density of
\nh$=$1--2$\times 10^{22}$cm$^{-2}$, a typical value of the cloud column
density. This corresponds to the peak of X-ray sources without a NIR counterpart. On the
contrary, a thin-thermal spectrum with \kt$=1.2$~keV, \Z$=$0.20, and
\nh$=10^{21}$cm$^{-2}$, typical values for class III and MS stars (see the following
section), has the hardness ratio of $\sim-$0.8. This corresponds to the peak of the
histogram of X-ray sources with a NIR counterpart. 

Therefore, X-ray sources with a NIR counterpart are mostly cloud members, consistent with
the conclusion of the previous subsection. We thus focus on the X-ray sources with a NIR
counterpart. X-ray sources without a NIR counterpart, on the other hand, are probably
AGNs, which are not the main subject of this paper.

\placefigure{fg:f2}

\subsection{Classification}
YSOs are observationally classified into class 0 through class III based on their
spectral energy distribution (SED) from optical, NIR, mid infrared (MIR), and
sub-mm bands \citep{andre94}. Since no systematic catalog for wavelengths longer than
MIR is published in both of these clouds, we use the \jh/\hk\ color-color diagram of the
2MASS counterparts for the classification \citep{lada92,strom95}. 

\begin{figure}
 \figurenum{3}
 \epsscale{0.8}
 \plotone{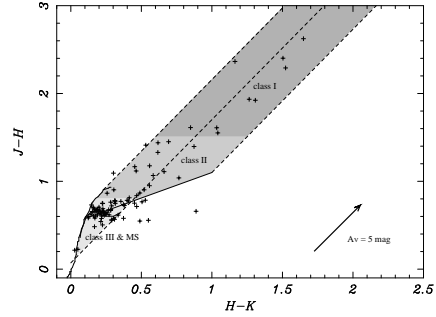}
 \caption{The \jh/\hk\ color-color diagram in CIT color system. ``Bright2'' sources with
 the 2MASS \ji, \hi, and \ksi-band detection are plotted (crosses). The typical error is
 $\pm$0.05 \magn. The evolution tracks of giants and dwarfs, and CTTS locus are given by
 solid lines, while the dotted lines represent the extinction vector. X-ray sources are
 classified into class~I, class~II, and class~III $+$ MS sources based on their position
 on the diagram.\label{fg:f3}}
\end{figure}

Out of the 123 ``bright2'' sources, 108 are detected either in the \ji, \hi, or
\ksi-band, and the remaining 15 are found in none of these bands. Figure~\ref{fg:f3}
shows the \jh\ vs \hk\ plots of the detected X-ray sources.

Intrinsic colors for giants and dwarfs are taken from \citet{tokunaga00} with colors
transformed into the CIT system \citep{bessel88}, while CTTS locus is taken from
\citet{meyer97}. We assume the slope of the reddening lines to be $E(\jh)/E(\hk)=1.63$
\citep{martin90}. All the 2MASS colors are also translated into CIT color system with
the transformation formula established by \citet{carpenter01}.

Based on the position in this diagram, we classify the sources into three groups ---
class~I, class~II, and class~III $+$ MS --- in the following way\footnote{Note that our
classification based only on three NIR bands may be less complete compared with the
conventional classification scheme. The number of class I and class II sources might be
underestimated because the J-H/H-K diagram is not very sensitive to the NIR excess
\citep{oloffson99,persi00,bontemps01}. On the other hand, the number of class I and
class II sources might be overestimated because we can not discriminate mildly
extinguished class III $+$ MS from them. Therefore it may be fair to use the terminology
``class I-like'' instead of ``class I'' etc. For simplicity, however, we use the latter
terminology in this paper.}.
Class I and class II sources are characterized by \hk\ color excess over \jh\ color,
originating from their disk emission, while class III and MS sources are not. Therefore,
sources between the second and the third leftmost reddening lines, and above the CTTS
locus are either class I or class II sources. Class I and class II sources in this
region can be distinguished from each other by the amount of extinction, where class Is
generally have higher extinction (typically \jh$>$1.5) than class
IIs~\citep{lada92,strom95}. Therefore, sources with \jh$>$1.5 are classified to be
class~I and \jh$\leq$ 1.5 are class~II. Sources between the leftmost and the second
leftmost reddening lines can be either class I, class II, class III or MS
stars. In this region again, we classify sources based on their amount of extinction;
sources with \jh$>$1.5 are class~I, \jh$>$0.8 are class~II, and \jh$\leq$0.8 are
class~III $+$ MS. These criteria are based on the fact that, in the Taurus-Auriga dark
cloud complex, class II sources rarely have larger extinction than \jh$>$1.5 and class
III sources rarely have larger extinction than \jh$>$0.8 \citep{strom95}. In addition,
sources which lack 2MASS \ji\ and/or \hi-band detection (the upper limit of the \hi-band
magnitude is given) are classified as class I, because all of them have larger
extinction of \hk$>$1.2 \magn.

Then, out of the 108 sources, we find 19 class~I, 18 class~II, and 61 class~III $+$ MS
sources. The other 10 are outside of the classification regions. The results are
summarized in Table~\ref{tb:t2}.

\placefigure{fg:f3}

\subsection{X-ray Properties of Different Classes}
Based on the classification described in the previous section, we list the X-ray
properties (absorption, metallicity, luminosity, and temperature) of each class in
Table~\ref{tb:t4} and compare them with each other.

\placetable{tb:t4}

\subsubsection{Absorption}
The X-ray absorption decreases as a YSO evolves from class~I through class~III $+$ MS. This
is basically consistent with the fact that we classify these sources using their NIR
extinction. \jh\ derived from the NIR photometry is a good indicator for the amount of
material in solid-state, while \nh\ derived from X-ray spectroscopy represents the
amount of gas. Therefore, \nh\ should go in proportion to \jh, and the ratio gives us
the information of the dust-to-gas ratio in a star forming cloud. 
 
\begin{figure}
 \figurenum{4}
 \epsscale{0.8}
 \plotone{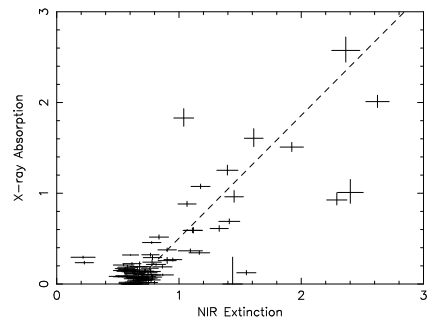}
 \caption{The NIR extinction (\jh\ \magn) and the X-ray absorption (\nh\
 10$^{22}$cm$^{-2}$) are plotted. The best-fit linear function is expressed with the
 dashed line.\label{fg:f4}}
\end{figure}

For all the ``bright2'' sources with NIR data, we plot the relation between \nh\ and
\jh\ (Figure~\ref{fg:f4}). A clear correlation is found. The best-fit linear function is
\begin{equation}
 \nh / 10^{22}~\rm{cm}^{-2} = (1.35\pm0.18)\{(\jh)-(0.63\pm0.05)\}\ \magn.
  \label{eq:abs-ext}
\end{equation}

\placefigure{fg:f4}

The \jh\ offset of 0.63$\pm$0.05 \magn\ should be the averaged intrinsic color after
removing the reddening, which corresponds to the spectral type K5--K7 in MS stars
\citep{tokunaga00}. This indicates that low mass YSOs and MS stars are dominant in these
clouds. The slope of 1.35$\pm$0.18 gives us information on the dust-to-gas
ratio. Together with the relation between $A_{V}$ and \jh\ given in Meyer et
al. (1997), 
\begin{equation}
 \nh / 10^{21}~\rm{cm}^{-2} = (1.49\pm0.20) \times A_{V}\ \magn
\end{equation}
is derived. The slope of 1.49$\pm$0.20 is smaller than those of the Galactic
interstellar medium and the $\rho$-Ophiuchi dark cloud, and is similar to that of another
star forming region, the Mon~R2 cloud \citep{predehl95,imanishi01,kohno01}. Thus the
dust-to-gas ratio may scatter from cloud to cloud, possibly with more massive star 
forming regions having larger dust-to-gas ratio.

\subsubsection{Metallicity}
We determine the metallicity for many sample stars in a star forming region for the first
time. A hint of a decreasing trend toward evolved classes is seen, although this depends on
the way the samples are broken into class~I, class~II, and class~III $+$ MS. The
metallicity, when all classes are combined, is 0.45 (0.37--0.52) solar. 

\citet{padgett96} observed 30 G and K pre-main-sequence stars in the nearby star forming
regions including Orion, and studied their photospheric abundances using the iron
absorption lines in the optical band. All star forming regions are found to have the
solar abundance.

The discrepancy between the coronal abundance and the photospheric abundance, which are
derived from the X-ray observations and the optical observations respectively, is often
seen in other samples \citet{yamauchi96,imanishi01}.

\subsubsection{Luminosity and Temperature}
\begin{figure}
 \figurenum{5}
 \epsscale{0.8}
 \plotone{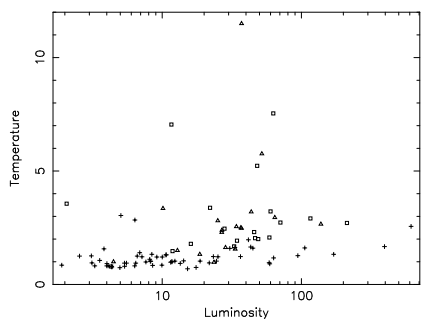}
 \caption{Luminosity (10$^{29}$ erg~s$^{-1}$) in logarithmic scale and temperature
 (keV) of each source are plotted separately for class~I (squares), class~II
(triangles), and class~III $+$ MS (crosses) sources.\label{fg:f5}}
\end{figure}

The temperature (\kt) and the luminosity (\lx) are plotted on Figure~\ref{fg:f5},
separately for class~I, class~II, and class~III $+$ MS. The temperatures of
class~I to class~III $+$ MS sources are randomly distributed over a wide range of
luminosity, but the mean temperatures of class~I and class~II sources are significantly
higher (\kt\ $\sim$ 3.0~keV) than that of class~III $+$ MS (\kt\ $\sim$ 1.2~keV). We may
argue that there are two types of X-ray emission mechanisms; one exhibits higher
temperature, and the other has lower temperature. The former dominates in less evolved
YSOs, like class I and class~II sources, while the latter gradually appears as YSOs
evolve to class~III $+$ MS. In this context, it is suggestive that most sources with two
temperature components (Table~\ref{tb:t4}) belong to class~III $+$ MS. The higher
temperature component of these sources has $\sim$ 2.3~keV on average similar to the 
1-temperature sources of class~I and class~II, while the mean of the lower temperature component 
is $\sim$ 0.8~keV similar to the 1-temperature sources of class~III $+$ MS.
We hence suspect that the two component class~III $+$ MS sources may be in the transition 
phase between higher-temperature-dominant and the lower-temperature-dominant stage.

\placefigure{fg:f5}

\subsubsection{Time Variation}
Class~II sources have a slightly higher fraction of time-variable sources than other
classes, though it is not statistically significant. When class~I and class~II are combined
and are compared with class~III $+$ MS, we see no significant difference in
time-variation rate (in both groups, $\sim$ 40\% are time-variable). \citet{imanishi01}
also argues that no significant difference of the flare rate is seen among three
classes in the $\rho$-Ophiuchi dark cloud. Whether time-variable activity is related
to the higher-temperature or the lower-temperature-component is not clear in our data set.

\section{SUMMARY}
We observed  OMC-2 and OMC-3 with the CXO/ACIS for $\sim$ 100~ksec. This is one of the
deepest observations ever performed in star forming regions in the X-ray band. We
detected $\sim$ 400 X-ray sources in our FOV. Coherent analyses on these sources derived
the following results.

\begin{enumerate}
 \item Imaging analysis is performed for all the detected sources, and their position,
       photon counts, and hardness ratio are derived. This is one of the largest catalogs
       of X-ray sources in a star forming region.
 \item Correlations with the 2MASS sources are found using the 2MASS database. About 60\% of
       the X-ray sources have a NIR counterpart. 
 \item Spectral and timing analysis are performed for ``bright2'' X-ray sources. A one
       temperature thin-thermal plasma model can explain most of the spectra. About
       40\% of the ``bright2'' sources are found to be time-variable. 
 \item Most of the X-ray sources with a NIR counterpart are likely cloud members, while
       those with no NIR counterpart are probably background AGNs.
 \item We classify the cloud members to be class I, class II, and class III $+$ MS based on
       the \jh/\hk\ color-color diagram and conduct a systematic comparison on the X-ray
       properties, such as absorption, luminosity, temperature, and time-variation.
 \item Class~I and class~II sources are found to have higher temperatures than
       class~III $+$ MS. We thus propose that two types of X-ray emission mechanisms
       exist (higher temperature and lower temperature component). The higher
       temperature plasma appears in the earlier phase and lower temperature plasma
       appears as YSOs evolve. In the transition phase, possibly early class III,
       plasma emissions with different temperatures coexist.
 \item The ratio of time-variable sources is nearly the same among different classes. 
       Whether the time-variability is related to higher-temperature- or
       lower-temperature-plasma is still an open question.
\end{enumerate}

\acknowledgments
The authors express their thanks to Dr. Yoshitomo Maeda for giving us invaluable
information on the analysis of the ACIS data, especially on the source detection
algorithm and detector calibrations. The authors also acknowledge the careful reading of
the manuscript by Jun Yokogawa and Leisa Townsley. M.T. and M.G. are financially supported by the Japan
Society for the Promotion of Science.


\clearpage
\begin{deluxetable}{lcrrrr}
 \tabletypesize{\scriptsize}
 \tablecaption{Results of Imaging Analysis and NIR Identification\tablenotemark{a}\label{tb:t1}}
 \tablecolumns{6}
 \tablewidth{0pt}
 \tablehead{
 \colhead{ID\tablenotemark{b}} & 
 \colhead{Name\tablenotemark{c}} &
 \colhead{X-ray\tablenotemark{d}} &
 \colhead{Hardness\tablenotemark{e}} & 
 \colhead{\jh\tablenotemark{f}} & 
 \colhead{\hk\tablenotemark{f}}\\
 \colhead{} & 
 \colhead{} & 
 \colhead{Counts} &
 \colhead{Ratio} & 
 \colhead{(\magn)} & 
 \colhead{(\magn)}
 }
\startdata
  I1 &  053438.60-050844.6 &    255  &   0.31  &  \nodata  &  \nodata\\	
  I2 &  053440.86-050700.0 &     63  &  -0.02  &  \nodata  &  \nodata\\	
  I3 &  053443.10-050919.8 &    368  &   0.62  &  \nodata  &  \nodata\\	
  I4 &  053444.89-050650.9 &    403  &  -0.54  &  0.725    &  0.379  \\	
  I5 &  053445.19-051046.9 &    176  &  -0.25  &  0.748    &  0.517  \\	
  I6 &  053448.29-050714.9 &    535  &  -0.66  &  0.658    &  0.223  \\	
  I7 &  053448.37-050502.9 &    549  &  -0.77  &  0.675    &  0.188  \\	
  I8 &  053449.18-050440.0 &    127  &  -0.97  &  0.695    &  0.227  \\
\enddata
 \tablenotetext{a}{Results of X-ray imaging analysis and 2MASS correlation are listed for
 all the detected X-ray sources.}
 \tablenotetext{b}{I1--I354 and S1--S11 are detected in the total band image
 (0.5--8.0~keV) image of the ACIS-I and the ACIS-S2. I355--I369 and S12--S13 are
 detected only in the hard band (2.0--8.0~keV) image of the ACIS-I and the
 ACIS-S2. I370--I385 are detected only in the soft band (0.5--2.0~keV) image of the
 ACIS-I. No new source was detected in the soft band image of the ACIS-S2.}
 \tablenotetext{c}{Sources are named after their position in the equinox J2000}
 \tablenotetext{d}{X-ray photon counts in 0.5--8.0~keV band.}
 \tablenotetext{e}{Defined as $H-S/H+S$ where $H$ and $S$ are photon counts in the hard
 band (2.0--8.0~keV) and in the soft band (0.5--2.0~keV) respectively.}
 \tablenotetext{f}{\jh\ and \hk\ in CIT color system derived from 2MASS \ji, \hi, and
 \ksi-band photometry and \citet{carpenter01}.}
\end{deluxetable}

\begin{deluxetable}{lcccccc}
 \tabletypesize{\scriptsize}
 \tablecaption{Results of Timing and Spectroscopic Analysis\tablenotemark{a}\label{tb:t2}}
 \tablecolumns{7}
 \tablewidth{0pt}
 \tablehead{
 \colhead{ID\tablenotemark{b}} & 
 \colhead{Upper\tablenotemark{c}} &
 \colhead{\nh \tablenotemark{d}} &
 \colhead{\kt \tablenotemark{de}} &
 \colhead{Metallicity \tablenotemark{deg}} & 
 \colhead{\lx \tablenotemark{df}} &
 \colhead{Class \tablenotemark{h}}\\
 \colhead{} &
 \colhead{Probability} &
 \colhead{($10^{22}$ cm$^{-2}$)} &
 \colhead{(keV)} &
 \colhead{(solar)} & 
 \colhead{(10$^{29}$ erg~s$^{-1}$)} &
 \colhead{}
 }
 \startdata
I4\udagger & 0.005 & 0.14 (0.12--0.16) & 1.40 (1.29--1.60) & 0.10 (0.07--0.12) & 6.90 (6.50--7.31) & III$+$MS \\
I6\udagger & 0.000 & 0.07 (0.05--0.08) & 0.99 (0.94--1.04) & 0.10 (0.09--0.12) & 7.64 (7.28--8.01) & III$+$MS \\
I7 & 0.014 & 0.14 (0.13--0.16) & 0.85 (0.81--0.88) & 0.07 (0.06--0.08) & 9.95 (9.50--10.41) & III$+$MS \\
I10 & 0.008 & 0.07 (0.05--0.08) & 0.98 (0.94--1.02) & 0.13 (0.12--0.14) & 11.51 (11.07--11.96) & III$+$MS \\
I12 & 0.196 & 0.14 (0.12--0.16) & 1.25 (1.16--1.34) & 0.13 (0.11--0.15) & 6.62 (6.24--6.99) & III$+$MS \\
I13\udagger & 0.000 & 0.07 (0.06--0.07) & 1.09 (1.06--1.13) & 0.03 (0.02--0.03) & 27.49 (26.78--28.19) & other \\
I14 & 0.079 & 0.02 (0.01--0.04) & 1.11 (1.06--1.34) & 0.08 (0.07--0.10) & 8.14 (7.78--8.50) & III$+$MS \\
I17 & 0.245 & 0.05 (0.04--0.07) & 1.02 (0.96--1.07) & 0.07 (0.06--0.08) & 8.28 (7.91--8.66) & III$+$MS \\
\enddata
 \tablenotetext{a}{Results of timing and spectroscopic analysis are listed for all the
 ``bright2'' sources. Best-fit parameters of thin-thermal plasma model fitting are
 tabulated regardless whether the fit is accepted or rejected.}
 \tablenotetext{b}{Sources with time-variability are shown with \udagger.}
 \tablenotetext{c}{Defined as $\int _{\alpha} ^{\infty} \chis _{\nu} (x) dx$ where the
 \chis -value and the degree of freedom of the fit is $\alpha$ and $\nu$. Fits with the
 upper probability larger than 0.01 are ``accepted'' while those less than or equal to
 0.01 are ``rejected''. Even the ``best-fit'' parameters of rejected fittings represent
 values of some physical meaning or approximated values. We thus use these values in the
 same way with the accepted ones.}
 \tablenotetext{d}{The lower and upper boundary for 1 $\sigma$ confidence level are
 given in parentheses.}
 \tablenotetext{e}{The best fit value of temperature and metallicity are searched for in
 the range of less than 20~keV and 5 solar respectively.}
 \tablenotetext{f}{Luminosity in 0.5--8.0~keV band (corrected for absorption).}
 \tablenotetext{g}{Abundance relative to solar photospheric \citep{morrison83}.}
 \tablenotetext{h}{Classification based on the \ji, \hi, and \ksi-band magnitude of
 2MASS counterpart. ``I'' for class~I, ``II'' for class~II, and ``III$+$MS'' for
 class~III$+$MS. ``others'' are for sources which are out of these groups. ``\nodata''
 sources can not be classified due to their lack of a NIR counterpart.}
\end{deluxetable}

\begin{deluxetable}{lccccccc}
 \tabletypesize{\scriptsize}
 \tablecaption{Best-fit Values of Two-Temperature Thin-Thermal Plasma Model\tablenotemark{a}\label{tb:t3}}
 \tablecolumns{8}
 \tablewidth{0pt}
 \tablehead{
 \colhead{ID} & 
 \colhead{\nh \tablenotemark{b}}&
 \colhead{\kt $_{1}$\tablenotemark{bce}}&
 \colhead{\kt $_{2}$\tablenotemark{bde}}&
 \colhead{Metallicity\tablenotemark{beg}}& 
 \colhead{\lx $_{1}$\tablenotemark{bcf}} &
 \colhead{\lx $_{2}$\tablenotemark{bdf}} &
 \colhead{Class \tablenotemark{g}}\\
 \colhead{} &
 \colhead{($10^{22}$ cm$^{-2}$)} &
 \colhead{(keV)} &
 \colhead{(keV)} &
 \colhead{(solar)} & 
 \colhead{(10$^{29}$ erg~s$^{-1}$)} &
 \colhead{(10$^{29}$ erg~s$^{-1}$)} &
 \colhead{}
 }
 \startdata
I10 & 0.42(0.41--0.44) & 0.22(0.21--0.22) & 1.07(1.02--1.12) & 0.43(0.41--0.45) & 33.08(30.39--35.88) & 12.31(11.62--13.00) & III$+$MS \\
I42 & 0.36(0.34--0.38) & 1.14(1.11--1.23) & 2.41(2.20--2.65) & 0.18(0.16--0.21)  & 12.43(11.55--13.29) & 17.62(16.61--18.64) & II \\
I47 & 0.36(0.35--0.38) & 0.79(0.75--0.83) & 2.46(2.31--2.62) &  0.47(0.45--0.50) & 15.30(14.28--16.32) & 36.59(35.36--37.82) & III$+$MS \\
I71 & 0.00(0.00--0.02) & 0.80(0.77--0.83) & 2.08(1.92--2.25) &  0.78(0.75--0.82) & 6.64(6.28--7.00) & 9.18(8.68--9.69) & III$+$MS \\
I150 & 0.24(0.23--0.25) & 0.83(0.81--0.85) & 10.51(6.34--26.52) & 0.33(0.32--0.34) & 39.42(38.43--40.41) & 12.26(11.11--13.43) & other \\
I156 & 0.22(0.19--0.25) & 0.84(0.72--0.98) & 2.53(2.26--9.32) & 0.61(0.52--0.70) & 1.39(1.11--1.67) & 4.62(4.21--5.02) & other \\
I174 & 0.00(0.00--0.01) & 1.06(1.04--1.09) & 3.71(3.01--4.69) & 0.29(0.27--0.30) & 17.93(17.37--18.48) & 9.80(8.99--10.62) & III$+$MS \\
I189 & 0.00(0.00--0.01) & 0.59(0.52--0.65) & 1.70(1.61--1.79) & 0.55(0.51--0.59) & 2.51(2.23--2.79) & 8.04(7.64--8.44) & III$+$MS \\
\enddata
 \tablenotetext{a}{Accepted fits are listed.}
 \tablenotetext{b}{The lower and upper boundary for 1 $\sigma$ confidence level are given in parentheses.}
 \tablenotetext{c}{Best-fit value of the lower temperature component.}
 \tablenotetext{d}{Best-fit value of the higher temperature component.}
 \tablenotetext{e}{The absorption and the metallicity of the higher temperature
 component is fixed to be the same with those of the lower temperature component. The
 best fit value of temperature and metallicity are searched for in the range of less
 than 20~keV and 5 solar respectively.}
 \tablenotetext{f}{Luminosity of each component in 0.5--8.0~keV band (corrected for absorption)}
 \tablenotetext{g}{Abundance relative to solar photospheric \citep{morrison83}.}
 \tablenotetext{h}{Classification based on the \ji, \hi, and \ksi-band magnitude of
 2MASS counterpart. ``I'' for class~I, ``II'' for class~II, and ``III$+$MS'' for
 class~III$+$MS. ``others'' are for sources which are out of these groups. ``\nodata''
 sources can not be classified due to their lack of a NIR counterpart.}
\end{deluxetable}

\begin{deluxetable}{cccc}
 \tabletypesize{\scriptsize}
 \tablecaption{X-ray Properties of Each Class\label{tb:t4}}
 \tablecolumns{4}
 \tablewidth{0pt}
 \tablehead{
 \colhead{} & 
 \colhead{Class I} &
 \colhead{Class II} &
 \colhead{Class III$+$MS}
 }
 \startdata
 Frequency \dotfill & 19 & 18 & 61\\
 Sources with time-variation \dotfill & 6 (32\%) & 10 (56\%) & 21 (34\%)\\
 Source with two temperature \dotfill & 1 ( 5\%) &  3 (17\%) & 15 (25\%)\\
 \hline
 Absorption(10$^{22}$cm$^{-2}$)\tablenotemark{a}\dotfill & 2.49(2.37--2.62) & 0.63(0.60--0.67) & 0.11(0.10--0.13)\\
 Metallicity (solar)\tablenotemark{ab}\dotfill & 0.82(0.61--0.99) & 0.60(0.50--0.69) & 0.20(0.18--0.22)\\
 Temperature (keV)\tablenotemark{a}\dotfill & 2.99(2.66--3.57) & 2.91(2.63--3.26)& 1.20(1.13--1.28)\\
 Luminosity (10$^{29}$erg~s$^{-1}$)\tablenotemark{a}\dotfill & 49.1(46.7--51.5) & 36.3(35.1--37.5) & 36.4(35.7--37.0)\\
 \enddata
 \tablenotetext{a}{The lower and upper boundary for 1 $\sigma$ confidence level are given in parentheses.}
 \tablenotetext{b}{Abundance relative to solar photospheric \citep{morrison83}.}
\end{deluxetable}


\clearpage
\begin{thebibliography}{}
 \bibitem[Andr\'{e} \& Montmerle(1994)]{andre94} Andr\'{e}, P. \& Montmerle, T. 1994,
					\apj, 420, 837
 \bibitem[Bessel \& Brett(1988)]{bessel88} Bessel, M. S. \& Brett, J. M. 1988, \pasp,
					100, 1134
 \bibitem[Bontemps et al.(2001)]{bontemps01} Bontemps, S. et al. 2001, \aap, 372, 173
 \bibitem[Carpenter(2001)]{carpenter01} Carpenter, J. M. 2001, \aj, 121, 2851
 \bibitem[Chini et al.(1997)]{chini97} Chini, R., Reipurth, B., Ward-Thompson, D.,
					Bally, J., Nyman, L-\AA, Sievers, A., \&
					Billawala, Y. 1997, \apj, 474, L135
 \bibitem[Feigelson \& DeCampli(1981)]{feigelson81} Feigelson, E. D. \& DeCampli 1981,
					\apj, 243, L89
 \bibitem[Garmire et al.(2001)]{garmire01} Garmire, G. P., et al., 2001, in preparation 
 \bibitem[Grosso et al.(2000)]{grosso00} Grosso, N., Montmerle, T., Bontempts, S.,
					Andr\'{e}, P., \& Feigelson, E. D. \aap, 359,
					113
 \bibitem[Imanishi et al.(2001)]{imanishi01} Imanishi, K., Koyama, K., \& Tsuboi,
				       Y. 2001, \apj, in press
 \bibitem[Kamata et al.(1997)]{kamata97} Kamata, Y., Koyama, K., Tsuboi, Y., \&
					Yamauchi, S. 1997, \pasj, 49, 461
 \bibitem[Kohno, Koyama, \& Hamaguchi(2001)]{kohno01} Kohno, M., Koyama, K., \&
				       Hamaguchi, K. 2001, in preparation
 \bibitem[Koyama et al.(1996)]{koyama96} Koyama, K., Hamaguchi, K., Ueno, S., \&
					Kobayashi, N. 1996, \pasj, 48, L87
 \bibitem[Krishnamurthi et al.(2001)]{krishnamurthi01} Krishnamurthi, A., Reynolds,
					C. S., Linsky, J.L., Martin, E., \& Gagn\'{e},
					M. 2001, \aj, 11, 337
 \bibitem[Lada \& Adams(1992)]{lada92} Lada, C. J. \& Adams, F. C. 1992, \apj, 393, 278
 \bibitem[Martin \& Whittet(1990)]{martin90} Martin, P. G. \& Whittet, D. C. B. 1990,
					\apj, 357, 113
 \bibitem[Meyer, Calvet, \& Hillenbrand(1997)]{meyer97} Meyer, M. R., Calvet, N., \&
					Hillenbrand, L. A. 1997, \aj, 114, 288
 \bibitem[Montmerle et al.(1983)]{montmerle83} Montmerle, T., Koch-Miramond, L.,
					Falgarone, E., \& Grindlay, J, E. 1983, \apj,
					269, 182
 \bibitem[Montmerle et al.(2000)]{montmerle00} Montmerle, T., Grosso, N., Tsuboi, Y.,
					\& Koyama, K. 2000, \apj, 532, 1097
 \bibitem[Morrison \& McCammon(1983)]{morrison83} Morrison, R. \& McCammon, D. 1983,
				       \apj, 270, 119
 \bibitem[Olofsson et al.(1999)]{oloffson99} Oloffson, G. et al. 1999, \aap, 350, 883
 \bibitem[Padgett (1996)]{padgett96} Padgett, D. L. 1996, \apj, 471, 847
 \bibitem[Persi et al.(2000)]{persi00} Persi, P. et al. 2000, \aap, 357, 219
 \bibitem[Predehl \& Schmitt (1995)]{predehl95} Predehl, P \& Schmitt, J. H. M. M. 1995,
				       \aap, 293, 889
 \bibitem[Schulz et al.(2001)]{schulz01} Schulz, N. S., Canizares, C., Huenemoerder,
					D., Kastner, J. H., Tayler, S. C., \& Bergstorm,
					E. J. \apj, 549, 441
 \bibitem[Skinner \& Walter(1998)]{skinner98} Skinner, S. L. \& Walter, F. M. 1998,
					\apj, 509, 761
 \bibitem[Strom, Kepner, \& Strom(1995)]{strom95} Strom, K. M., Kepner, J., \& Strom,
					S. E. 1995, \apj, 438, 813 
 \bibitem[Tokunaga(2000)]{tokunaga00} Tokunaga, A. T. 2000, in Allen's Astrophysical
				       Quantities, ed. A. N. Cox, (4th ed.; New York:
				       Springer-Verlag), 143
 \bibitem[Tsuboi et al.(2000)]{tsuboi00} Tsuboi, Y., Imanishi, K., Koyama, K., Grosso,
					N., \& Montmerle, T. 2000, \apj, 532, 1089
 \bibitem[Tsuboi et al.(2001)]{tsuboi01} Tsuboi, Y., Koyama, K., Hamaguchi, K.,
					Tatematsu, K., Sekimoto, Y., Bally, J., \&
					Reipurth, B. 2001, \apj, 554, in press
 \bibitem[Weisskopf, O'dell, \& van Speybroeck(1996)]{weisskopf96} Weisskopf, M. C.,
					O'dell, S., \& van Speybroeck, L. P. 1996,
					Proc. of SPIE, 2805, 2
 \bibitem[Yamauchi et al.(1996)]{yamauchi96} Yamauchi, S., Koyama, K., Sakano, M., \&
					Okada, K. 1996, \pasj, 48, 719
\end{thebibliography}
\end{document}